\documentclass[twocolumn,twoside,slac_two]{revtex4}
\usepackage{graphicx}
\usepackage{fancyhdr}
\begin{document}

\title{Equation of State and the Finite Temperature Transition in QCD}

%

\author{Rajan Gupta [HotQCD Collaboration]}
\affiliation{Theoretical Division, Los Alamos National Laboratory, Los Alamos, NM 87545, USA}

\begin{abstract}
This talk provides a summary of the results obtained by the HotQCD
collaboration on the equation of state and the crossover transition in
2+1 flavor QCD. We investigate bulk thermodynamic quantities - energy
density, pressure, entropy density, and the speed of sound over the
temperature range $140 < T < 540$ MeV.  These results have been
obtained on lattices of temporal size $N_\tau = 6$ and $8$ and with
two improved staggered fermion actions, asqtad and p4.  Our most
extensive results are with masses of the two degenerate light quarks
set at $m_{l} = 0.1 m_s$ corresponding to the Goldstone pion mass
$m_\pi$ between $220-260 MeV$. In these simulations, the strange quark
mass is tuned to its physical value and constant values of $m_l/m_s$
define lines of constant physics. We also summarize the current state
of results on observables sensitive to the chiral and deconfining
physics $-$ the light and strange quark number susceptibilities, the
chiral condensate and its susceptibility, and the renormalized
Polyakov loop. Our results indicate that the deconfinement and chiral
symmetry restoration occur in the same narrow temperature interval.
\end{abstract}

\maketitle

\thispagestyle{fancy}


\section{Introduction}

Ongoing experiments at RHIC and proposed experiments at LHC aim to
understand the properties of hot dense nuclear matter created in the
collision of two relativisitic nuclei. At sufficiently high
temperatures and densities RHIC data support the creation of a
quark-gluon plasma in the central region that undergoes a transition
back to hadronic matter as it expands and cools. The goal is to
explain the creation and evoluton of this medium. Hydrodynamic
descriptions used to model this evolution provide a good fit to the
data and are thus the phenomenological tool of choice.  First
principle calculations using lattice QCD yield a number of properties
of QCD as a function of temperature that are essential inputs in these
hydrodynamical analyses.  These properties include the nature of the
transition (with respect to both confinement and chiral symmetry
breaking) between the quark-gluon plasma and hadronic matter, the
transition temperature, the equation of state of QCD and transport
coefficients as a function of temperature in the range $140-700$ MeV.

HotQCD is a US wide collaboration engaged in the study of QCD at
finite temperature and density using lattice QCD
\footnote{HotQCD Collaboration members are: 
A. Bazavov, T. Bhattacharya, M. Cheng, N.H. Christ, C. DeTar, S. Ejiri, 
S. Gottlieb, R. Gupta, U.M. Heller, K. Huebner, C. Jung, F. Karsch, 
E. Laermann, L. Levkova, C. Miao, R.D. Mawhinney, P. Petreczky, C. Schmidt, 
R.A. Soltz, W. Soeldner, R. Sugar, D. Toussaint and P. Vranas}. 
It brought together
members of the MILC and RBC-Bielefeld collaborations to carry out
large scale simulations on IBM Bluegene/L supercomputers at the
Lawrence Livermore National Lab and on the NYBlue at the New York Center for 
Computational Sciences at BNL.

Our goals are to perform detailed simulations of $2+1$ flavor QCD
using staggered fermions at the physical values of the strange and
light quark masses. Most of the results presented here are for $
m_l = 0.1 m_s$, about a factor of two heavier than the mean physical
$u$ and $d$ quark mass, $m_l \approx 0.038 m_s$. Improvements in
algorithms and computer resources are being used to incrementally
approach the physical up and down quark masses and the continuum
limit, thus providing high precision results. Also, in the current
simulations the up and down quark masses are taken to be degenerate.

The physical quantities we are calculating include:
\begin{itemize}
\item The Equation of State (EoS) of QCD over the temperature range
  $140-700$ MeV that is being probed by relativistic heavy ion
  experiments at Brookhaven and will be studied in more detail in the
  future at the LHC.

\item The nature of the deconfinement transition between the hadronic
  phase at low temperature and the quark-gluon plasma at high
  temperatures. All simulations with staggered fermions show a rapid
  crossover rather than a genuine phase transition for physical values
  of the light quark masses. In the absence of a phase transition, there is {\it a priori} no
  unique transition temperature as it can depend on the probe. Thus,
  the temperature at which this transition takes place remains a
  subject of investigations. The status of current estimates is
  discussed at the end of this paper. 
\item The restoration of chiral symmetry at high temperature and
  whether this chiral transition is coincident with the deconfining
  transition.

\item A detailed understanding of the physics in the transition region
  and the approach of thermodynamics quantities such as pressure, entropy, 
  energy density and the speed of sound, to the Stefan-Boltzmann limit. 
\end{itemize}

To understand and control systematic errors in the lattice calculations 
we are taking the following steps: 

\begin{itemize}
\item Simulations at $N_\tau = 4, 6, $ and $8$ have been performed
  with two improved staggered fermion actions -- asqtad and p4 -- to
  understand $O(a^2)$ discretization errors.

\item Extrapolation to the continuum limit to remove discretization
  errors will be made using simulations at $N_\tau = 6, 8$ and a new
  set of $N_\tau =12$ lattices.

\item The theory possesses a chiral phase transition in the limit
  $m_l = 0$.  To study the chiral behavior, extrapolations to the
  physical light quark mass limit and the chiral limit will be made
  using simulations at $m_l / m_s = 0.2,\ 0.1$ and $0.05$.

\item The transition region, $160-230$ MeV, is being sampled in
  intervals of 5 MeV or less using high statistics simulations in
  order to pin down the position of the peak in the chiral
  susceptibility and the inflection point in the light quark number
  susceptibility.
\end{itemize}

The first part of this report summarizes results for the equation of
state obtained by the HotQCD collaboration~\cite{hotqcd09}. In the
second part we discuss our results for the temperature dependence of
the chiral and deconfining transitions and finally compare these
results against recent data presented by the Wuppertal-Budapest
collaboration~\cite{aoki09}.

\section{Setup of Lattice Calculations}

To determine the lattice parameters for a systematic finite
temperature analysis, zero temperature simulations are used to first
set the lattice scale $a$ and the physical value of the strange quark mass
as a function of $\beta$.  The scale $a$ is extracted using the Sommer scale
$r_0$ obtained from the derivative of the static $q \bar q$ potential
evaluated on the lattice
\begin{equation}
\left( r^2 \frac{dV_{q \bar q}}{dr} \right)_{r=r_0} = 1.65
\end{equation}
with $r_0 = 0.469(7)$ fm taken from the Upsilon $2S-1S$
splitting~\cite{wingate2004}.  The strange quark mass is then set
using the $s \bar s$ pseudoscalar meson state with $M_{ss} r_0 = 1.58$.

The last quantity to fix are the light (up and down) quark masses that
are taken to be degenerate. We define lines of constant physics (LCP)
by keeping $m_l / m_s$ fixed.  In our calculations three values
of $m_l / m_s = 0.2,\ 0.1$ and $0.05$ are being
simulated. Results for the physical value of the light quark mass are
then obtained by extrapolation in $m_l $ at fixed lattice
scale. With these definitions, the continuum limit along a LCP can be
taken by varying a single paramter, the gauge coupling $\beta \to
\infty$ or equivalently $a \to 0$ since $\beta$ and $a$ are related by
dimensional transmutation in QCD.  The results presented here are
obtained mostly for $m_l / m_s = 0.1$ with $N_\tau = 6$ and $8$
lattices, so estimates of extrapolated values and uncertainties in
them should be considered preliminary.

In the finite temperature calculations, the temperature $T$, lattice
scale $a$ and the size of the lattice in the temporal direction
$N_\tau$ are related as $ T = 1/a N_\tau$. For fixed $N_\tau$ we
increase $T$ by decreasing $a$ or equivalently increasing $\beta$, and adjusting $m_s$ for each $\beta$.

There are two overarching issues with the use of staggered
fermions. Both have to do with the fact that staggered fermions
preserve a residual chiral symmetry at the expense of a 4-fold
doubling of flavors, i.e., every continuum quark flavor develops four
copies on the lattice. These four copies are called taste. The first
issue with using staggered quarks is one of principal. To simulate a
single flavor (say an s quark) we take the fourth root of the fermion
determinant in calculating the fermion force in Monte Carlo
simulations. Creutz~\cite{creutz07} has argued that a strong mixing of
tastes with different chiralities leads to an incorrect 't Hooft
vertex, and concludes that rooting can often be a good approximation
but predictions for non-perturbative physics where the 't Hooft vertex
is important can not be trusted.  On the other hand
Sharpe~\cite{sharpe06} and Goltermann~\cite{golterman08} argue that
while this rooting trick is ugly a well defined continuum limit exists
and the theory recovered in this limit is QCD.  Numerical results
obtained with staggered fermions when compared to experimental data
indicate that the rooting trick works~\cite{durr05} $i.e.$ this
circumstantial evidence provide support that staggered simulations
reproduce QCD. Clearly, our calculations rely on the validity of the
arguments reviewed by Sharpe and Golterman. In short, our work does not shed any
independent light on this issue.

The second issue is that, at finite lattice spacings, the 4 copies
called tastes, are not degenerate and the taste symmetry is badly
broken for both asqtad and p4 actions at $1/a \leq 2$ GeV
corresponding to the transition region for $N_\tau \leq 8$
simulations. This issue is being explored and improved versions of the
staggered action are being designed to reduce this breaking. The most
severe impact of this breaking is in the pion sector. In our p4 action
calculations with $m_l / m_s = 0.1$ on $N_\tau = 8$ lattices, the mass
of the Goldstone pion is $M_\pi r_0 \approx 0.52$ corresponding to
$M_\pi \approx 220$ MeV while the other fifteen higher mass taste
flavor pions are at $ 440$ MeV or heavier. The splitting for asqtad
action is slightly smaller. We expect this violation of taste symmetry
to affect the results in the low temperature confined phase and in the
transition region where the dominant excitations are
pions. Unfortunately, at this point we do not have a good estimate of
the uncertainty introduced in the various quantities we calculate due
to taste symmetry breaking.

The single basic quantity given by lattice simulations en-route to
calculating thermodynamics variables such as the pressure and energy
density is the trace anomaly ${\Theta^{\mu\mu} a^4} =
(\varepsilon-3p)a^4$.  We first discuss the data for it and the
extraction of the EoS from it. The second part introduces probes used
to understand the nature and location of the deconfining and chiral
symmetry restoration transitions and presents our data. Finally we
compare these results to those from the Wuppertal-Budapest
collaboration~\cite{aoki09}. 

\section{Trace Anomaly $(\varepsilon - 3 p)/T^4$}

The calculation of ${\Theta^{\mu\mu} a^4}$ involves calculating terms
that make up the gauge and fermion actions. As a result, most of the
computer time is spent in generating the ensemble of statistically
independent gauge configurations.  Results presented here represent
over 100 million node hours on the BlueGene/L.

\begin{figure}[h]
\centering
\includegraphics[width=80mm]{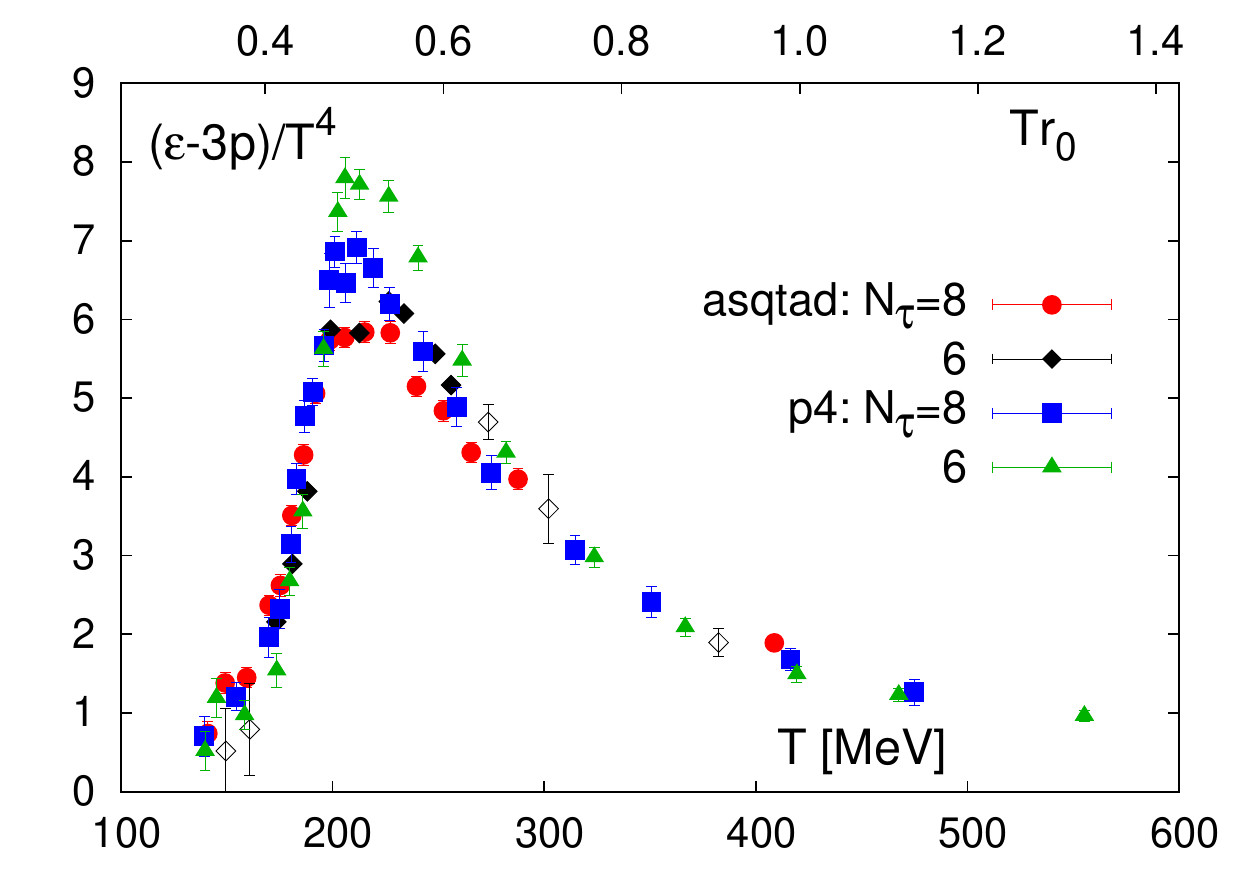}
\caption{Data for the trace anomaly $(\varepsilon-3p)/T^4$ calculated
  using the asqtad and p4 actions at $N_\tau=6,8$. This is the single
  basic quantity obtained from lattice simulations.}
\label{fig:e3p_full}
\end{figure}

\begin{figure}[ht]
\centering
\includegraphics[width=80mm]{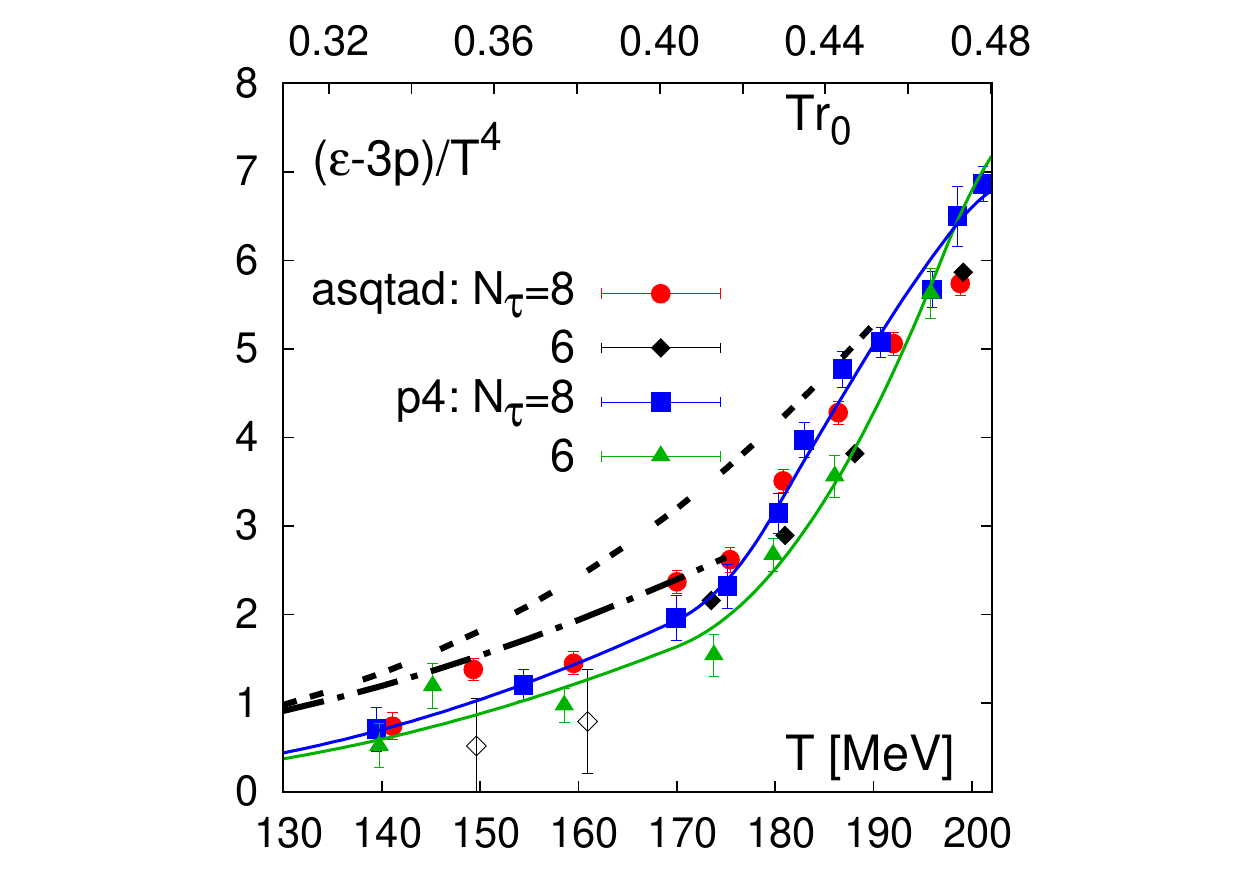}
\caption{Details of $(\varepsilon-3p)/T^4$ in the range $140-200$ MeV.} 
\label{fig:e3p_low}
\end{figure}

\begin{figure}[ht]
\centering
\includegraphics[width=80mm]{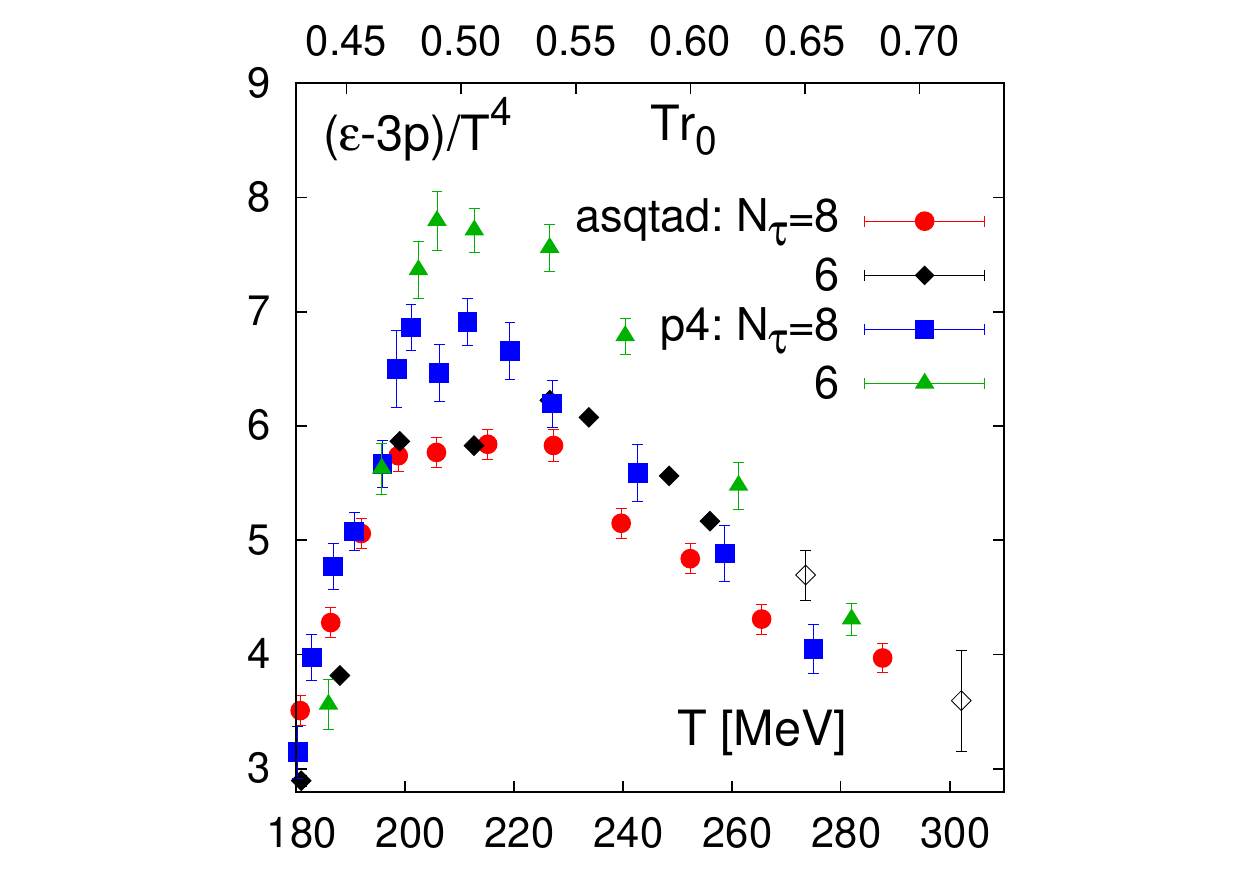}
\caption{Details of $(\varepsilon-3p)/T^4$ over the range $180-300$ MeV.} 
\label{fig:e3p_mid}
\end{figure}

\begin{figure}[ht]
\centering
\includegraphics[width=80mm]{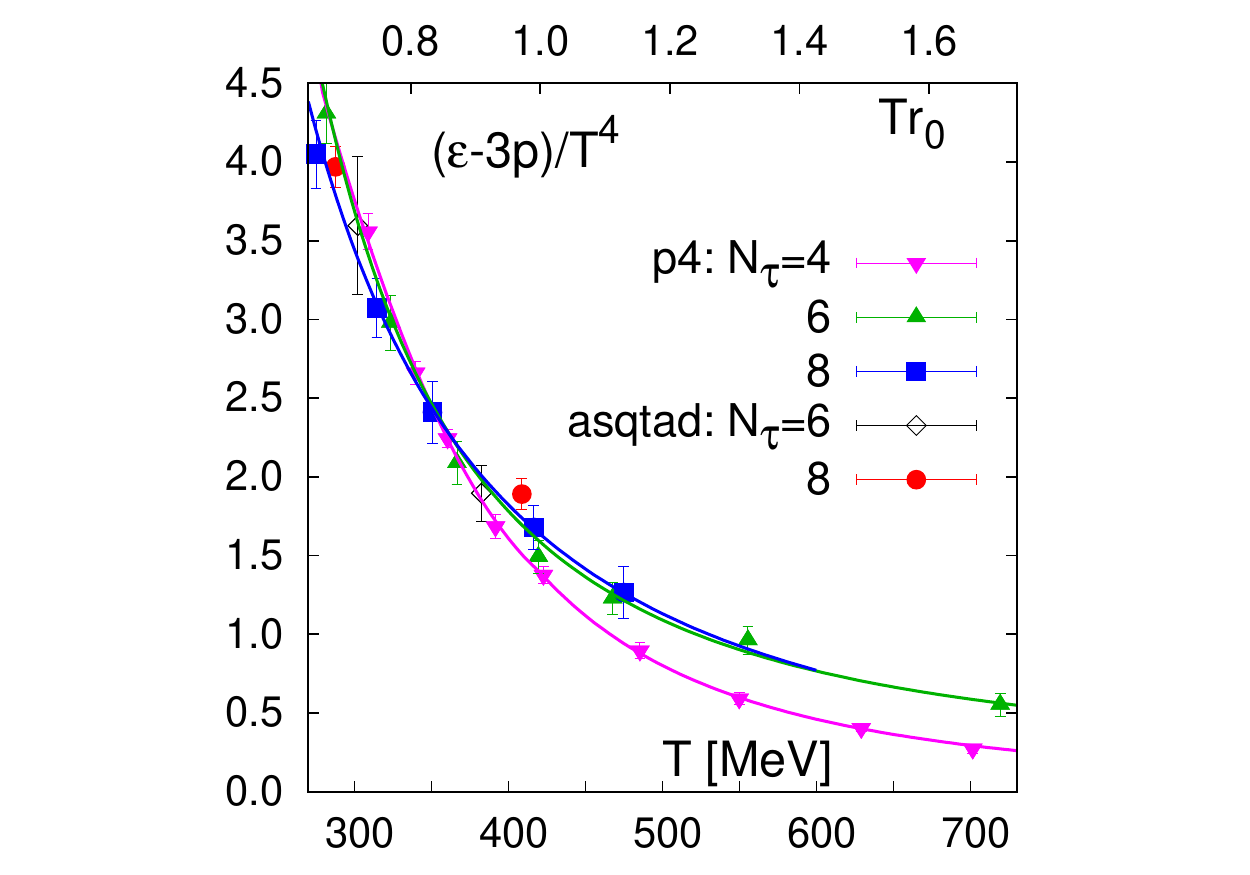}
\caption{Details of $(\varepsilon-3p)/T^4$ over the range $300-700$ MeV.} 
\label{fig:e3p_high}
\end{figure}

Data for $(\varepsilon-3p)a^4$ is converted into
physical units using the scale $r_0$ and subsequently in units of the
temperature $T$ as shown in Fig.~\ref{fig:e3p_full}. Overall, the
data show that the asqtad and p4 actions give consistent results, with
differences consistent with the expected magnitude of $O(a^2)$
errors. The details are highlighted in figures \ref{fig:e3p_low},
\ref{fig:e3p_mid}, \ref{fig:e3p_high} and discussed below.  

From the trace anomaly the pressure is obtained using the relations
\begin{eqnarray}
\frac{\Theta^{\mu\mu}}{T^4}   \equiv  \frac{\varepsilon - 3p} { T^4} & = & T \frac{\partial}{\partial T} \left( \frac{p}{T^4} \right)  \nonumber \\
\frac{p(T)}{T^4} - \frac{p(T_0)}{T_0^4}  & =  & \int_{T_0}^{T} dt \frac{\Theta^{\mu\mu}(t)}{t^5}
\label{eq:tanomaly}
\end{eqnarray}
The energy density, entropy density $(\varepsilon + p)/T^4$ and the
speed of sound are then given by appropriate combinations of
$(\varepsilon-3p)/T^4$and $p/T^4$.

{\bf Low Temperature Region:} Comparing $N_\tau=6$ and $8$ data in
Fig.~\ref{fig:e3p_low} for the low temperature region show a $\sim 5 $
MeV shift towards lower temperatures under $N_\tau=6 \to 8$. This
shift is indicative of $O(a^2)$ discretization errors. To calculate
the pressure using Eq.~\ref{eq:tanomaly} we need to fit the data
obtained at a discrete set of points to a functional form that can be
integrated and fix the normalization of \( \Theta^{\mu\mu} \) at some
low starting temperature $T_0$. We first discuss the normalization and
return to the determination of the functional form in the next
section.  For $T < 150$ MeV, the discretization errors in lattice
simulations, especially those due to the large taste symmetry
violations in asqtad and p4 staggered actions, are pronounced so
$N_\tau=8$ data have unknown systematic uncertainties. We, therefore,
explore predictions of the hadron resonance gas (HRG) model as it is
an independent method in which control over systematic uncertainties
improves as $T$ is taken below the pion mass scale $M_\pi = 140$
MeV~\cite{andronic06}. In Fig. \ref{fig:e3p_low} we plot results for
two choices of the cutoff scale $m_{max}$ up to which resonances are
included, $m_{max}=1.5$ GeV (dot dashed black curve) and $2.5$ GeV
(upper dashed curve).  Unfortunately, the HRG estimates show
significant sensitivity to the cutoff $m_{max}$ for $T > 100$ MeV.
Also, there is no matching point $T_0$ between lattice and the HRG
data.  There is a trend in the lattice results to move towards the HRG
values as $a \to 0$, i.e., as $N_\tau=6 \to 8$, however, at his point,
we do not have a way to quantify it, and conclude that further work is
required to reduce the uncertainty associated with the choice of
normalization.

{\bf Intermediate (Peak) Region:} Figure~\ref{fig:e3p_mid} shows that the peak
in ${\Theta^{\mu\mu}/T^4}$ occurs at $T > 200$ MeV. The differences
between the asqtad and p4 results for $N_\tau=6$ versus $ 8$ are the
largest in the interval $ 200 \leq T \leq 300$ and show a $15-20\%$
effect at the peak. We also find that the peak in the asqtad data is
shallower and shows smaller difference between $N_\tau=6$ and $8$
values as compared to the p4 results.

{\bf High temperature region $T > 300$ MeV:} Data in
Fig.~\ref{fig:e3p_high} show that the two actions give consistent
results and the difference between $N_\tau=6$ and $8$ are small.
${\Theta^{\mu\mu}/T^4}$ should, at sufficiently high temperatures,
vanish as $g^4(T)$, however, between $300-700$ MeV the data show a
much faster variation. We, therefore, use the ansatz
\begin{equation}
\frac{\varepsilon - 3p}{T^4} = \frac{3}{4} b_0 g^4 + \frac{d_2}{T^2} + \frac{d_4}{T^4}
\label{eq:highT_fit}
\end{equation}
to fit the data. Fits show no sensitivity to the leading $g^4$
term. Consequently, this term is neglected in the fits shown in
Fig.~\ref{fig:e3p_high} and used for the extraction of $p$.

To improve these fits and determine the shape of the curves for $T
>300$ MeV two enhancements to our data sets are needed. First, we need
data at more points. Second, checks for finite volume corrections need
to be made for $T > 500$ MeV when the lattice size in the spatial
directions used in the simulations, $N_s=32$, approaches $N_s a =
1/160$ for $a$ in MeV${}^{-1}$. Here $a $ corresponds to the zero
temperature scale at each of the the gauge couplings used to simulate
$T> 500$ MeV lattices, and the temperature scale $160$ MeV is used as
it marks the rapid onset of thermal fluctuations as shown in
Fig~\ref{fig:e3p_full}.

\subsection{Fits to ${\Theta^{\mu\mu}/T^4}$ and extraction of $p$}

\begin{figure*}[htb]
\centering
\includegraphics[width=80mm]{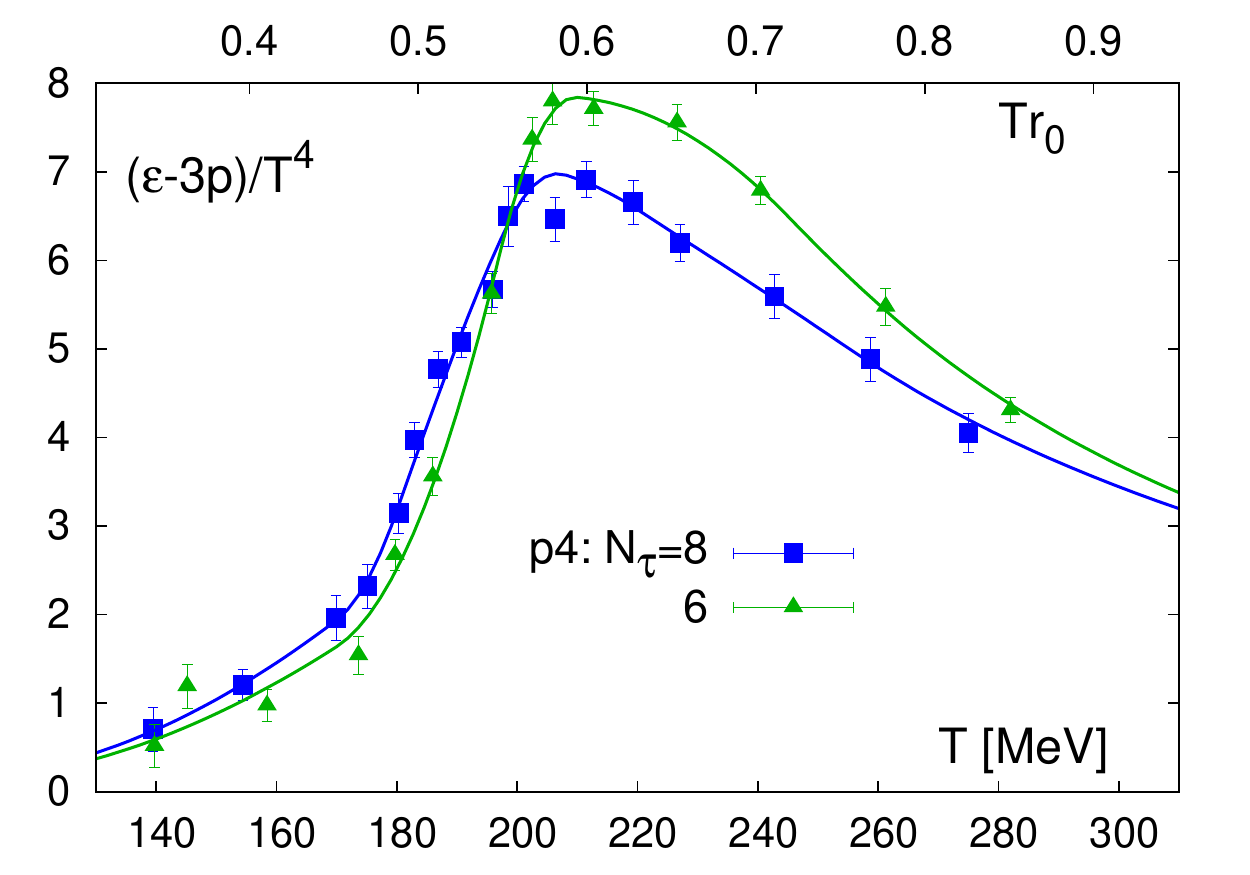}
\includegraphics[width=80mm]{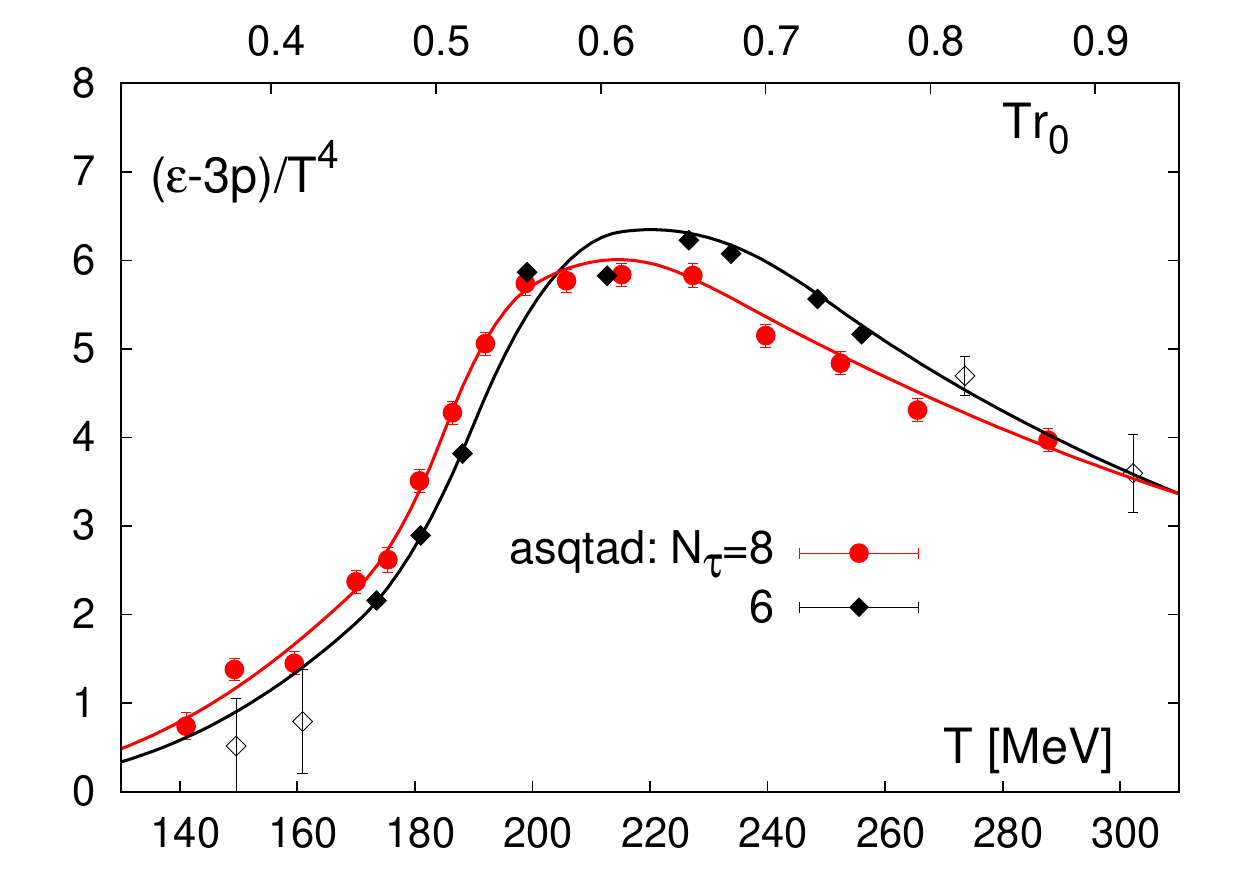}
\caption{Fits to ${\Theta^{\mu\mu}/T^4}$ data for the p4 lattice action  (left panel) and the asqtad action (right panel).} 
\label{fig:fit_interpolation}
\end{figure*}

%

To extract the behavior of $p$ versus $T$ using Eq.~\ref{eq:tanomaly}
we need to fit the data for ${\Theta^{\mu\mu}/T^4}$ calculated at a
discrete set of points with a smooth function. We do this using a
piecewise smooth interpolation. For $T < 170$ MeV an exponential fit
is used, while for $170 < T < 250$ MeV separate quadratic fits over
several intervals were used.  We found it necessary to match the
function and its derivative at each end of these intervals to get a
smooth curve for $p$. For $T> 250$ MeV the result of the fit with the
form given in Eq.~\ref{eq:highT_fit} is used. These fits are shown for
both actions and for $N_\tau = 6$ and $8$ in
Figures~\ref{fig:fit_interpolation}. 

The fits are very similar in shape for the two actions. In each case
shifting the fits to $N_\tau=6$ data by $\approx 5$ MeV to lower
temperatures give much better agreement except for the height of the
peak in the p4 data. This shift by $5$ MeV suggests the size of
discretization errors coming from a combination of (i) the errors in
the finite $T$ simulations and (ii) from the scale $a$ extracted from
zero temperature simulations clustered around $\beta$ values
corresponding to the two peak regions for $N_\tau=6$ and $8$.

\begin{figure}[ht]
\centering
\includegraphics[width=80mm]{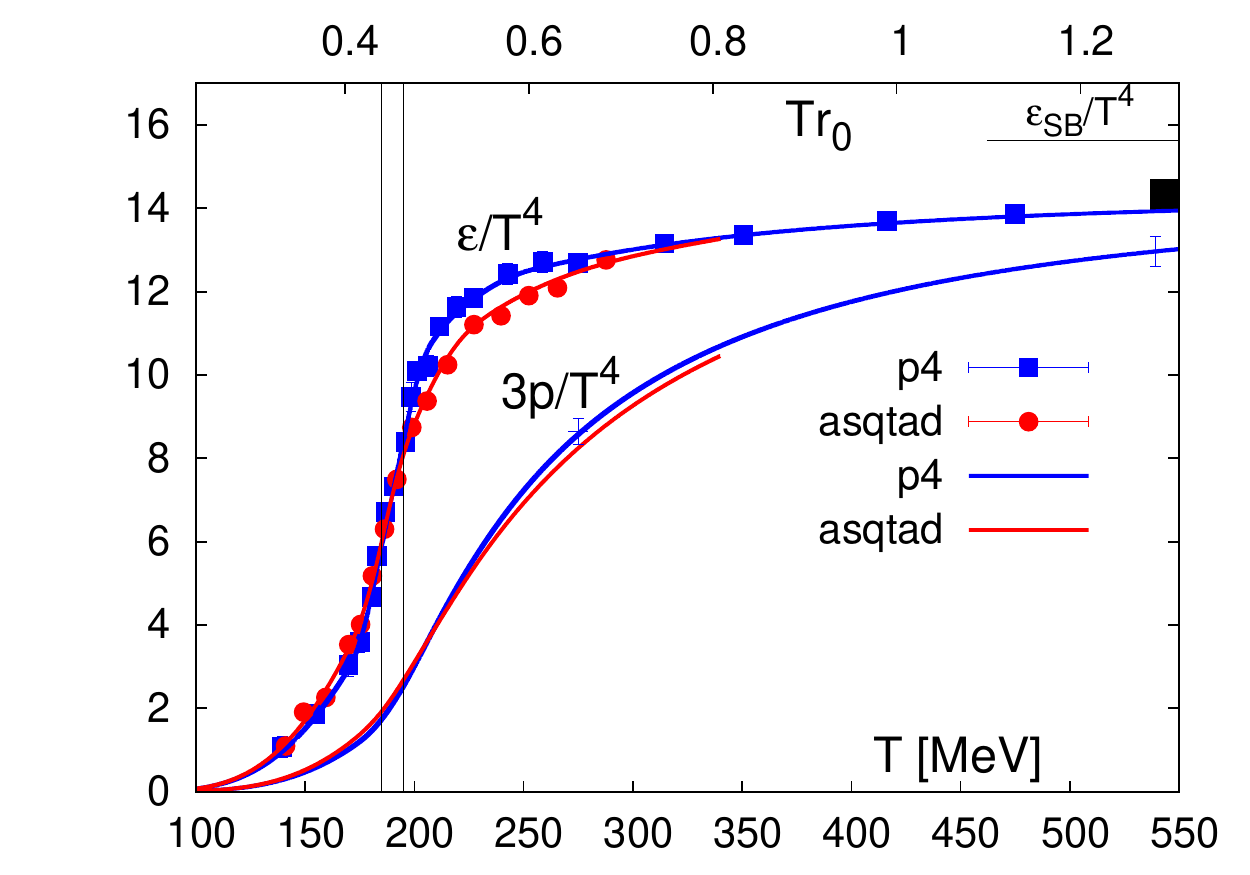}
\caption{Results for the energy density $\varepsilon/T^4$ and the pressure $3 p /T^4$ for the two actions 
obtained on $N_\tau=8$ lattice with $m / m_s = 0.1$.} 
\label{fig:PandE}
\end{figure}

Our results for $p$ using Eq.~\ref{eq:tanomaly} (and consequently all
other thermodynamic quantities) are obtained with the normalization
$p=0$ at $T=T_0=0$ and requiring the fit to the lattice data to pass
through this point. Redoing the fits starting with the HRG value at
$T=100$ MeV results in a global shift in the pressure and energy
density curves by $\approx 0.8$. The size of this effect is 
illustrated by the black square at $T=550$ MeV in
Fig.~\ref{fig:PandE}.

\begin{figure}[ht]
\centering
\includegraphics[width=80mm]{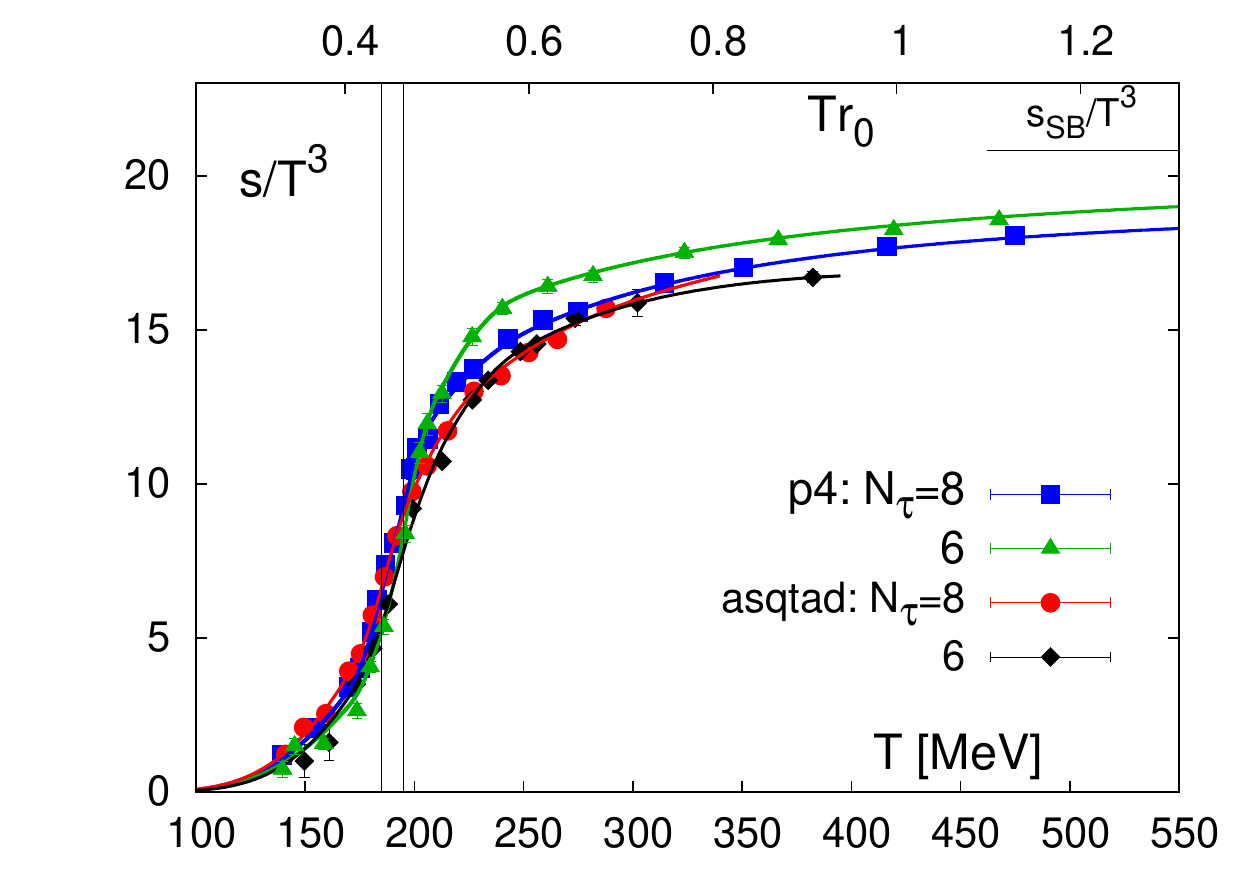}
\caption{The entropy density obtained on 
lattices with temporal extent $N_\tau =6$ \cite{milc_eos,rbcBIeos} and $8$.}
\label{fig:entropy}
\end{figure}

\begin{figure}[ht]
\centering
\includegraphics[width=80mm]{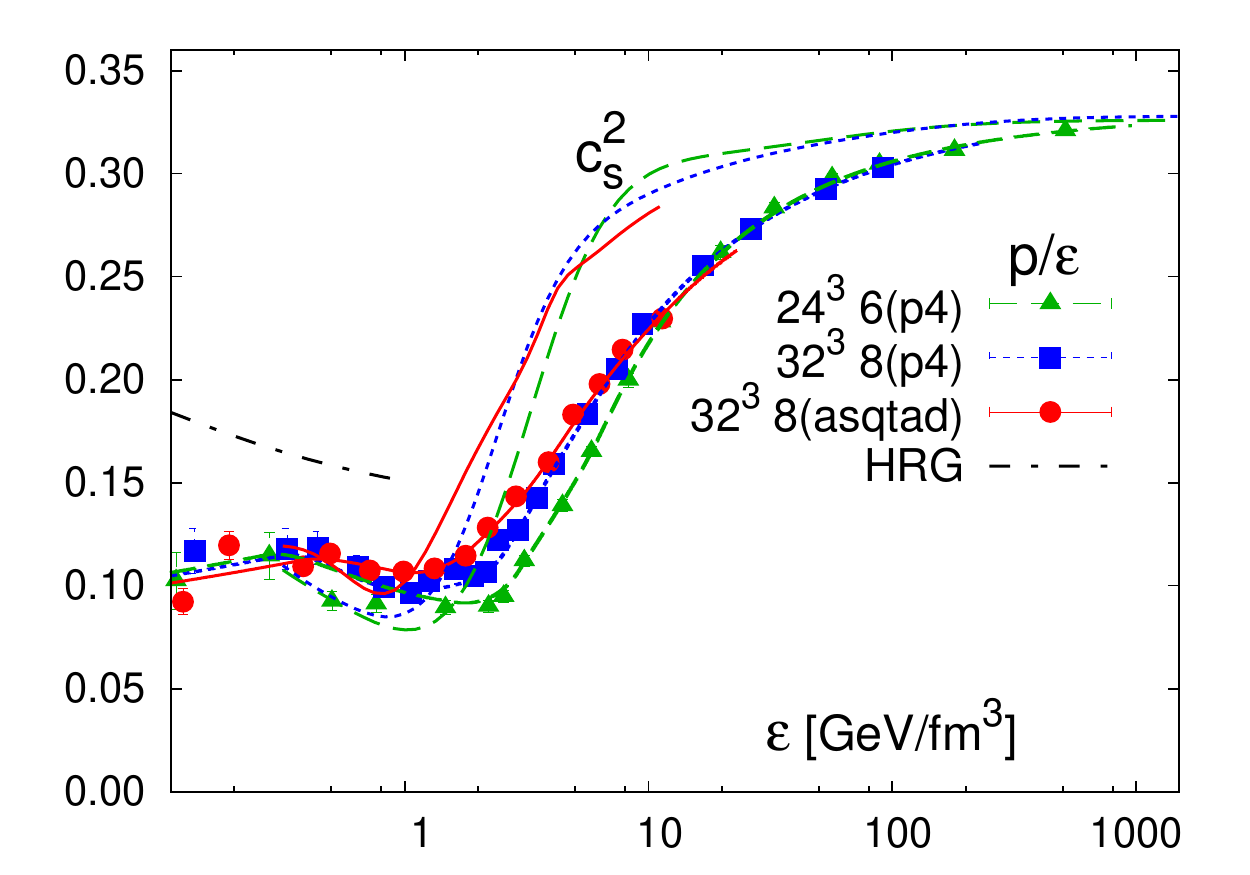}
\caption{Data for $p/\varepsilon$ and the speed of sound extracted
  from it. Lines without data points are obtained using the
  interpolating functions shown in
  Figures~\ref{fig:fit_interpolation}.  The dash-dot line is the HRG
  estimate.}
\label{fig:sound_speed}
\end{figure}

Once $({\varepsilon - 3p})/{T^4} $ and $p/T^4$ are known we get the
other thermodynamics quantities; entropy density as $({\varepsilon +
  p})/{T^4} $ and the speed of sound by using the ratio
$p/\varepsilon$ in
\begin{equation}
c_s^2 = \frac{{\rm d} p}{{\rm d}\epsilon} = \epsilon 
\frac{{\rm d} (p/\epsilon)}{{\rm d}\epsilon} + \frac{p}{\epsilon}\; .
\label{sound}
\end{equation}
These results are shown in Figures~\ref{fig:entropy} and
\ref{fig:sound_speed}. We regard the spread in the results between the
two actions as indicative of discretization errors.

In Figures~\ref{fig:PandE} and \ref{fig:entropy} we highlight the band
$185-195$ MeV which we estimate covers the inflection point in the
energy density $\varepsilon$ and the entropy density data.  Both of these
thermodynamics quantities probe deconfinement $-$ the change in the
thermodynamics as the degrees of freedom evolve from hadrons to quarks
and gluons due to the change in their free energies. Note that while
we discuss other probes of deconfinement and chiral symmetrey
restoration later, it is these thermodynamic quantities that are inputs
in the hydrodynamic analysis of RHIC data.

Our data show that for $T > 300$ MeV the degrees of freedom are quarks
and gluons and the deviations from the Stefan-Boltzmann limit are
$10-15\%$. Thus experiments at the LHC, which will probe the full
range $140 < T < 700$ MeV, will be able to explore the behavior and
nature of collective excitations in the quark-gluon plasma. Even
though the HotQCD estimates for the EoS do not include extrapolation
to the continuum limit and physical light quark masses, we propose
that they are realistic and ready for use in phenomenological
analyses.

\section{The Deconfining and Chiral Transition}

The data presented in previous sections show that the transition from
hadronic matter to quark-gluon plasma is a rapid crossover and not a
genuine phase transition.  This is consistent with all existing
calculations with staggered fermions.  We expect a singularity in the
partition function to exist in the chiral limit. What remains to be
settled for QCD with a physical value of the strange quark mass is
whether this phase transition occurs only at zero light quark masses
($m_l\equiv m_{crit}=0$) or at small but nonzero value ($m_l\equiv
m_{crit}>0$). In the first case the transition would belong to the
universality class of the 3-dimensional $O(4)$ symmetric spin model and in
the latter case to the $\rm 3-d$ Ising model~\cite{milc04}.

In light of this expected behavior, the questions that we would like
to answer regarding the nature and location of the transition from
hadronic matter to quark-gluon plasma in QCD are: 
\begin{itemize}
\item  How is the observed crossover in thermodynamic quantities for physical 
  light and strange quark masses influenced by the chiral phase transition?
\item What do standard probes of deconfinement (expectation value of
  Polyakov loop and quark number susceptibilities) and chiral
  symmetry restoration (chiral condensate and the chiral
  susceptibility) tell us?
\item Do the crossovers indicating deconfinement and chiral symmetry
  resoration happen at the same temperature (we will denote these as
  $T_c$ but note that it represents the crossover temperature and not
  a critical temperature) and what are the mechanisms that drive them?
\end{itemize}
Our results for the various probes are presented next.

\subsection{Renormalized Polyakov Loop}

\begin{figure}[ht]
\centering
\includegraphics[width=80mm]{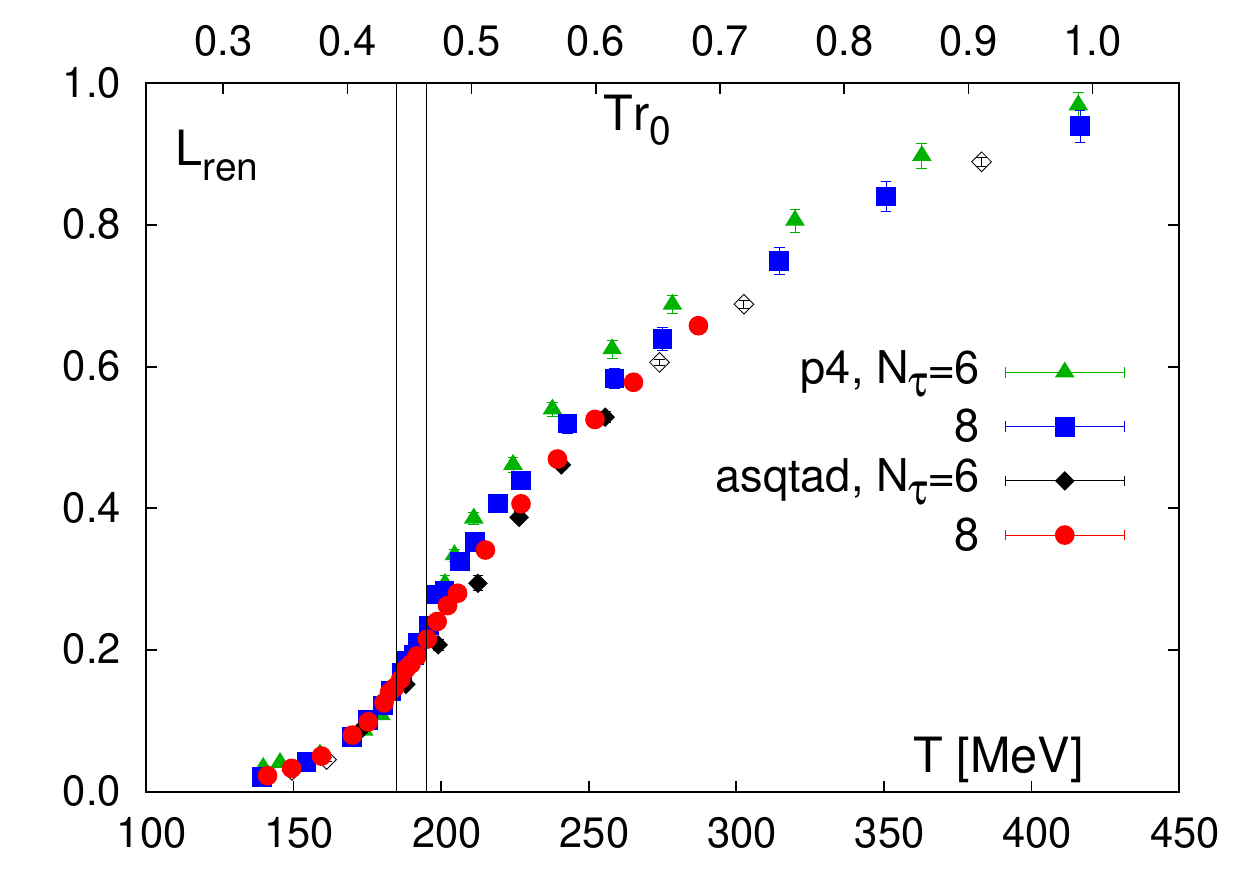}
\caption{Data for the Polyakov Loop $\langle L_{ren} \rangle $. Note the gradual rise 
in contrast to the rapid crossover seen in thermodynamic quantities. }
\label{fig:polyakov}
\end{figure}

The logarithm of the expectation value of the Polyakov loop is related
to the free energy of an isolated quark, 
\begin{equation}
\langle L \rangle \sim \exp{(- F_q / T)} \ .
\end{equation}
In QCD $\langle L \rangle$ is not a genuine order parameter, $i.e.$,
zero in the confined phase and non-zero in the deconfined phase, for
any finite values of the quark masses.  As far as we know, the
Polyakov loop is not directly sensitive to the singular structure of
the partition function in the chiral limit. It does , however, exhibit
a broad crossover as shown in Fig.~\ref{fig:polyakov} where we plot
the renormalized quantity $\langle L_{ren} \rangle = Z(\beta)^{N_\tau}
\ \langle L \rangle $~\cite{rbcBIeos}.  The renormalization is needed
to remove the self-energy contributions to the free energy.

Data for $\langle L_{ren} \rangle $ exhibit a gradual rise rather than
a rapid crossover. This feature is also reflected in the lack of a
peak in its susceptibility.  We find only a broad shoulder after the
rise for all data with $N_\tau \geq 6$ lattices. Since there is no
well defined inflection point in $\langle L_{ren} \rangle $ or a peak
in its susceptibility we do not consider it a reliable probe and argue
it should not be used in precision studies to determine the location
of a crossover $T_c$ indicative of deconfinement.

\subsection{Quark Number Susceptibility}

Thermal fluctuations of the degrees of freedom that carry a net number
of light or strange quarks are probed by taking derivatives of the
partition function with respect to the corresponding chemical potential, 
\begin{equation}
\frac{\chi_{q}}{T^2} = \frac{1}{VT^3} 
\left( \frac{\partial^2\ln Z}{\partial(\mu_{q}/T)^2} \right)_{\mu_q =0} \; ,\;\; q=l,\; s \ .
\label{chi_q}
\end{equation}

\begin{figure}[ht]
\centering
\includegraphics[width=80mm]{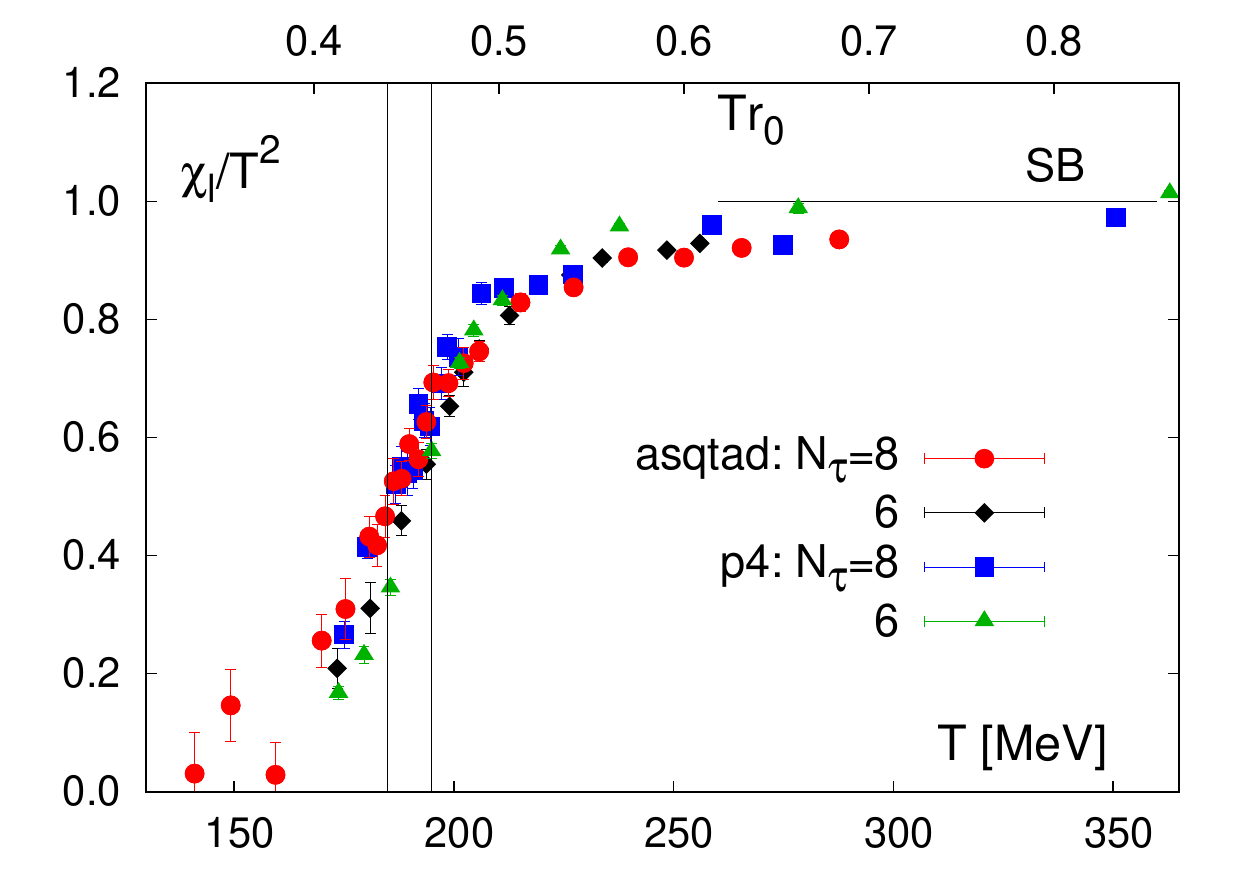}
\caption{Data for the light quark number susceptibility $\chi_l/T^2$. Note the 
crossover is consistent with the band at $185-195$ MeV and the value approaches 
unity rapidly after $250$ MeV. }
\label{fig:chi_qns_l}
\end{figure}

The quark number operator $\bar \psi \gamma_0 \psi $ is
a good probe in lattice calculations since it does not require
normalization. It counts the charge $N_q$, {\it i.e.}, the net number
of light or strange quarks. Similarly $\chi_q \sim \langle
N_q^2\rangle$. Therefore, in the continuum and infinite temperature limit
$\lim_{T\rightarrow\infty} \chi_q/T^2 = 1$, the Stefan-Boltzmann value
for an ideal massless one flavor quark gas. The approach to unity in
our data for the light quarks is shown in Fig.~\ref{fig:chi_qns_l}.

\begin{figure}[ht]
\centering
\includegraphics[width=80mm]{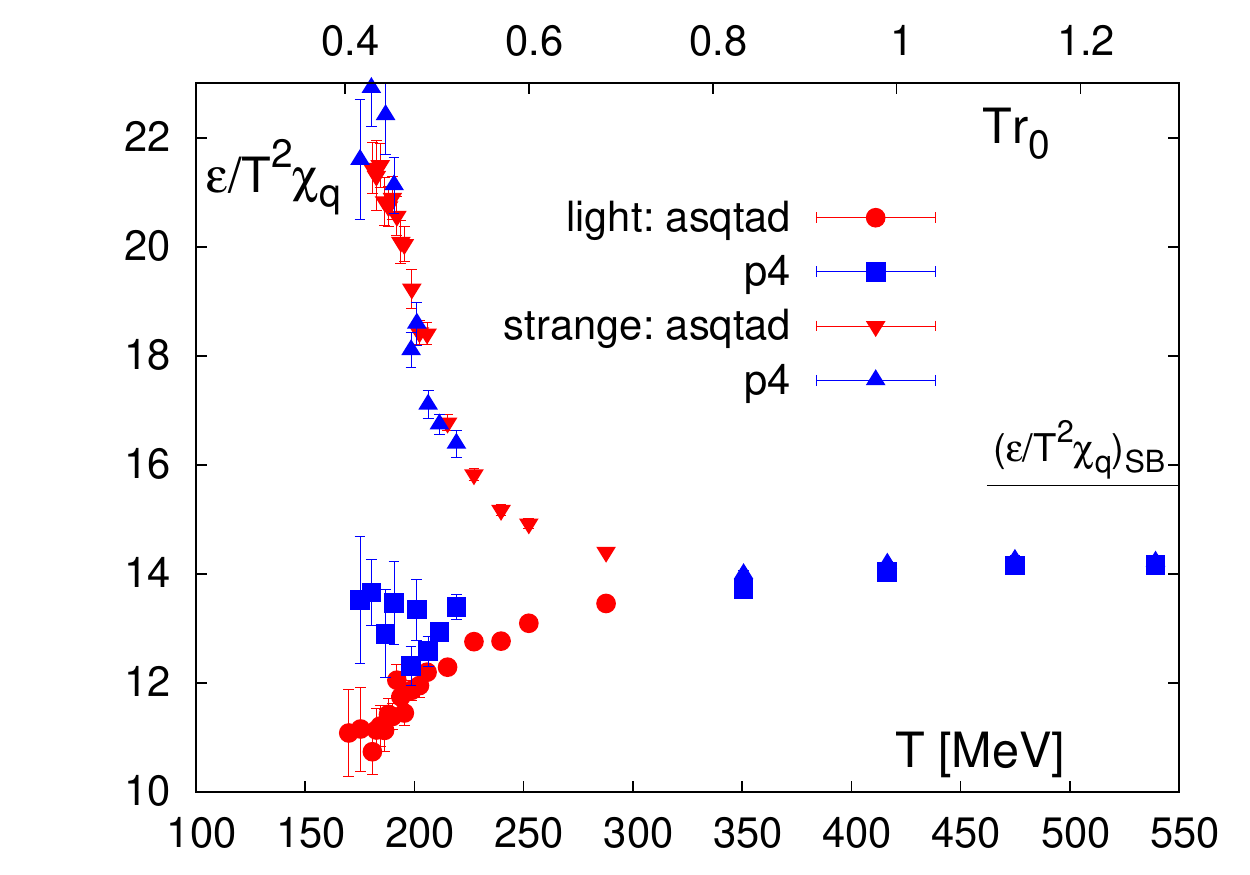}
\caption{Comparison of the strange and light quark number susceptibility in units of 
the energy density, $i.e.$ $\varepsilon / \chi_s T^2$ versus $\varepsilon / \chi_l T^2$. $\varepsilon/\chi_s T^2$ 
is seen to grow rapidly and is expected to diverge as $m_l \to 0$ since $\varepsilon$ gets contributions from 
the pions while $\chi_s \to 0$ as $\exp (-M_K/T)$. }
\label{fig:chi_qns_compare}
\end{figure}

As the transition region is approached from above, hadronic states
with appropriate quantum numbers start to contribute in proportion to
their Boltzmann weight $\sim \exp (-M/T)$.  At low temperatures the
dominant contributions are from the lightest hadronic state that has
the correct quantum numbers, $i.e.$, $\chi_s/T^2 \sim \exp (-M_K/T)$,
while $\chi_l/T^2 \sim \exp (-M_\pi/T)$.  Since $M_K$ remains finite
in the limit $m_l \to 0$, only $\chi_l/T^2$ is directly sensitive
to singularities in the QCD partition function in the chiral
limit. This expectation is confirmed by comparing the behavior of
$\varepsilon / \chi_s T^2$ versus $\varepsilon / \chi_l T^2$ in
Fig.~\ref{fig:chi_qns_compare}. Below the transition region
$\varepsilon / \chi_s T^2$ rises rapidly in contrast to $\varepsilon /
\chi_l T^2 $. In fact it will diverge because $\chi_s$ is not
sensitive to the lightest hadrons and goes to zero faster than
$\varepsilon$ which does receive contributions from the pions. We
therefore claim that $\chi_l/T^2 $ and not $\chi_s/T^2 $ should be
used to probe the singular structure of the theory.

The data for $\chi_l/T^2 $ in the transition region (see
Fig.~\ref{fig:chi_qns_l}) show a rapid crossover with the location of the
inflection point again consistent with the band $185-195$ MeV. We
therefore conclude that all probes of the deconfinement transition,
$\varepsilon$, the entropy density and $\chi_l/T^2 $, give consistent
results.

\subsection{Chiral Condensate and Susceptibility}

The light and strange quark chiral condensates are given by the
derivatives of the free energy with respect to the quark masses:
\begin{equation}
\langle \bar{\psi}\psi \rangle_q = \frac{T}{V}
\frac{\partial \ln Z}{\partial m_q} \; , \;\; q=l,\; s,
\label{eq:chiral}
\end{equation}
At finite values of the quark masses $\langle \bar{\psi}\psi
\rangle_q$ requires both multiplicative and additive
renormalizations. Both renormalizations are performed by defining the
following subtracted quantity for light quarks
\begin{equation}
\Delta_{l,s}(T) = \frac{\langle \bar{\psi}\psi \rangle_{l,T} -
\frac{m_l}{m_s}
\langle \bar{\psi}\psi \rangle_{s,T}}{\langle \bar{\psi}\psi \rangle_{l,0} -
\frac{m_l}{m_s} \langle \bar{\psi}\psi \rangle_{s,0}} \; .
\label{delta_ls}
\end{equation}
The subtractions proportional to $\langle \bar{\psi}\psi
\rangle_{s}$ remove the singular part of the additive
renormalization proportional to $m_q/a^2$. Division by the $T=0$
estimate obtained at the same lattice scale $a$ cancels the
multiplictive renormalization as it is $T$ independent.

\begin{figure}[ht]
\centering
\includegraphics[width=80mm]{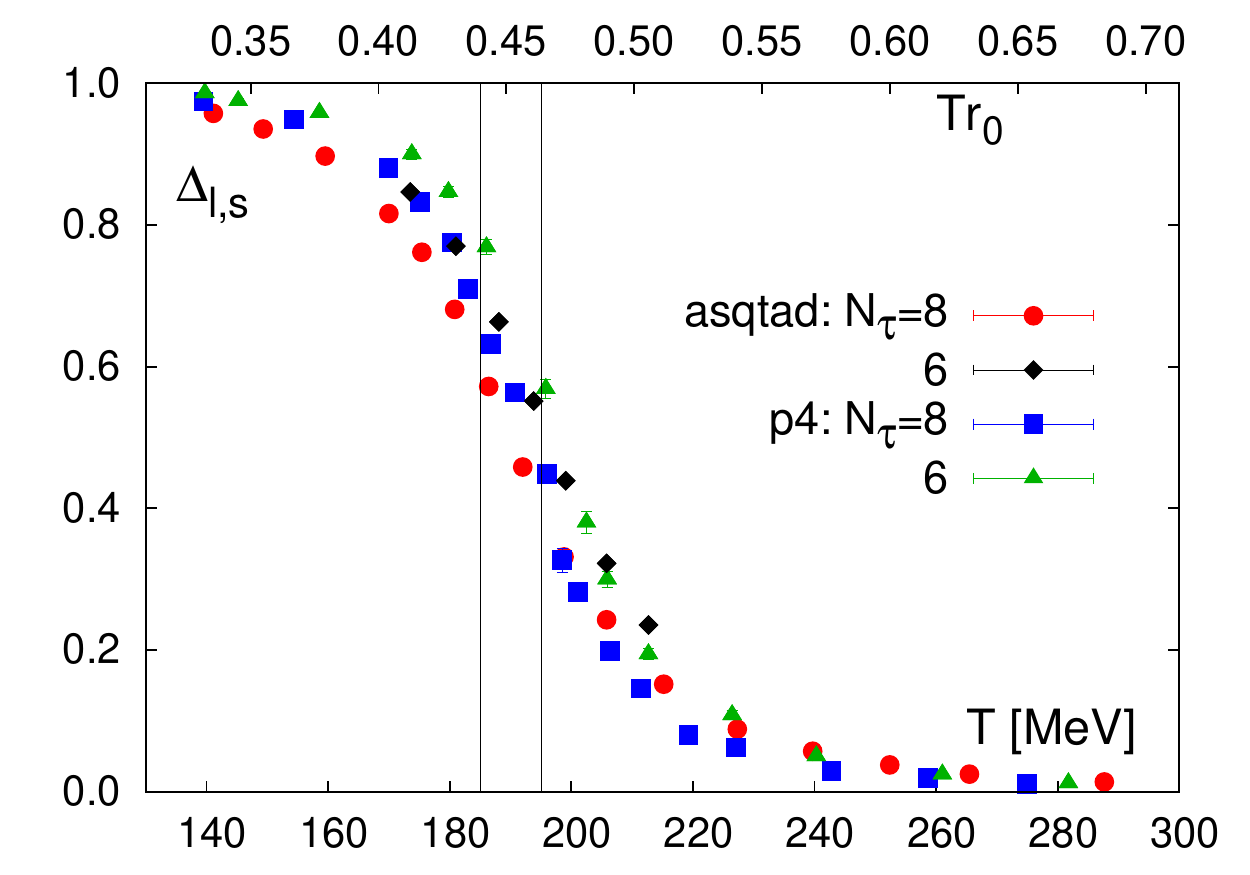}
\caption{p4 and asqtad data for the renormalized chiral condensate $\Delta_{l,s}(T)$ 
along the LCP given by $m_l=0.1 m_s$. }
\label{fig:condensate}
\end{figure}

Data for $\Delta_{l,s}(T)$ along the LCP given by $m_l=0.1 m_s$ are
shown in Fig.~\ref{fig:condensate}. We find a rapid crossover in the
same temperature range as exhibited by the bulk thermodynamic
observables and the light quark number susceptibility. This indicates
that deconfinement and chiral symmetry restoration happen at the same
temperature.

\begin{figure}[ht]
\centering
\includegraphics[width=80mm]{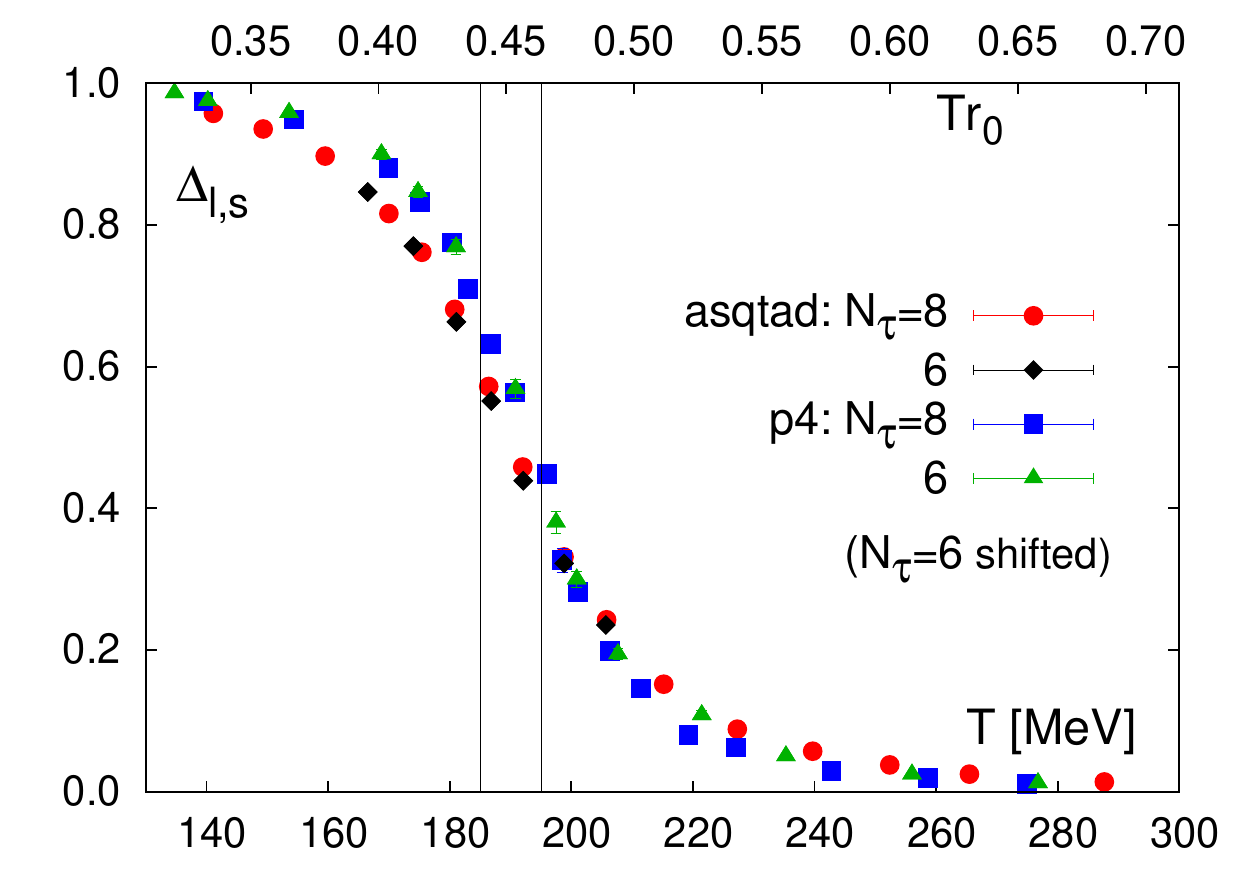}
\caption{Same data as in Fig.~\ref{fig:condensate} for the renormalized chiral condensate $\Delta_{l,s}(T)$ 
but with the $N_\tau=6$ data shifted to the left by $5$ MeV for each action. }
\label{fig:condensate_shifted}
\end{figure}

To estimate the magnitude of the difference between $N_\tau = 6$ and
$8$ data we plot, in Fig.~\ref{fig:condensate_shifted}, the same data
after shifting the $N_\tau=6$ values by 5 MeV to the left. The results
for $N_\tau=6$ and $8$ lattices now come together. The replot also
brings out the difference in results between the two actions $--$ the
difference is in the shape of curves. The data with the p4 action
shows a slightly steeper crossover.  Based on
Fig.~\ref{fig:condensate_shifted} we estimate that both systematic
effects are $\sim 5$ MeV and the location of the inflection point is
covered by the band $185-195$ MeV shown by the vertical lines.

The chiral susceptibility is given by the second derivative of $\ln Z$ with respect 
to the quark mass $m$:
\begin{eqnarray}
\frac{\chi_{l,s}}{T^2} &= &\frac{T}{V} \left( \frac{\partial^2\ln Z}{\partial^2 (m_{l.s}/T)^2} \right) \nonumber \\
      & = & \langle {\rm Tr}(M^{-2}_{l,s}) \rangle   \nonumber \\
      & & +\   \frac{1}{4} \left(  \langle ({\rm Tr} M^{-1}_{l,s})^2 \rangle  - \langle {\rm Tr} M^{-1}_{l,s} \rangle^2 \right) ,
\label{chi_sus}
\end{eqnarray}
where the first term in the right hand side gives the contributions of connected Feynman diagrams and the second gives 
those from the disconnected diagrams. 
$\chi_l$ is the best probe of the chiral transition as it exhibits a peak
even when the theory has a crossover and it will diverge at the chiral
phase transition, $i.e.$, in the chiral limit.  The location of the
peak is used to mark $T_c$.

\begin{figure*}[ht]
\centering
\includegraphics[width=77mm]{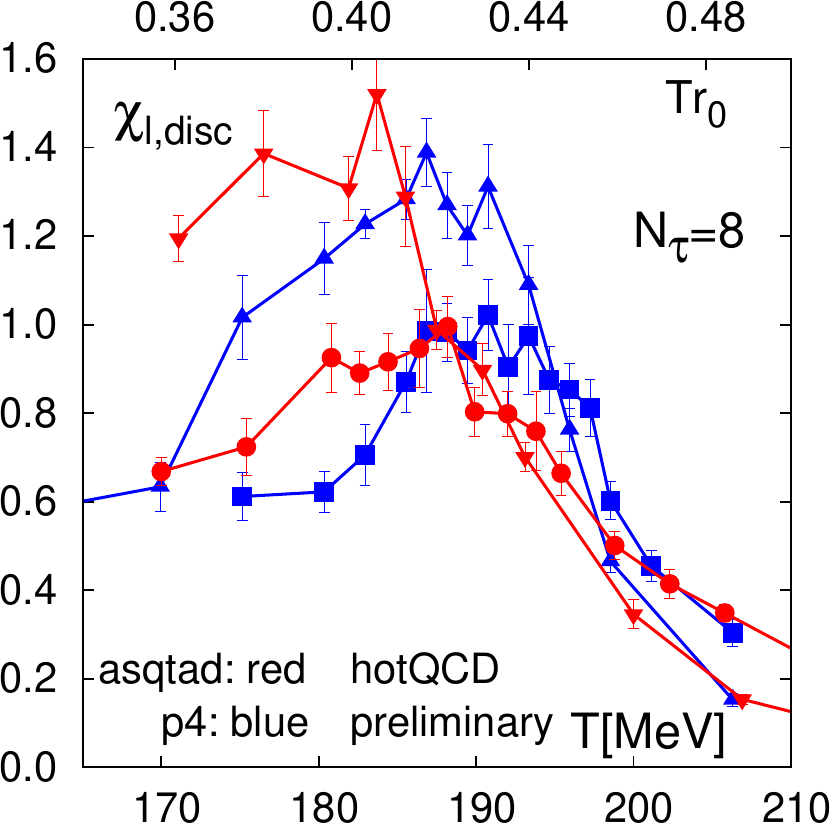}
\hspace{22pt}
\includegraphics[width=80mm]{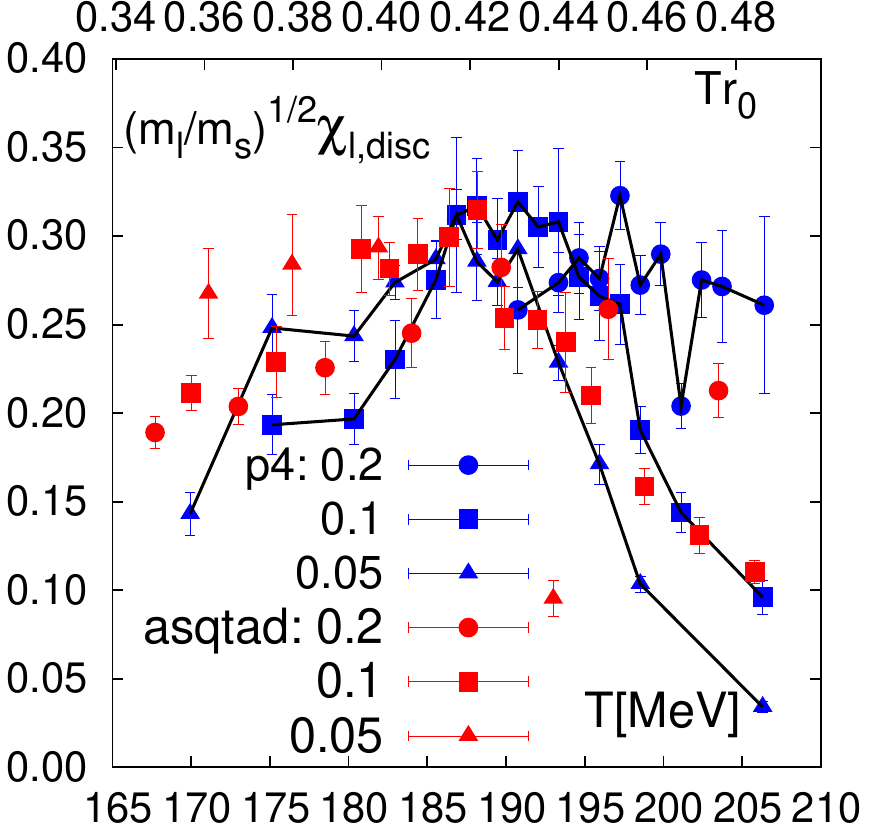}
\caption{Preliminary HotQCD data for the disconnected part of the
  chiral susceptibility $\chi_{l}(T)$ for the p4 and asqtad actions
  with $N_\tau=8$ and three values of the quark mass $m_l / m_s =
  0.2$, $0.1$ and $0.05$.  The right panel shows the same data as in the
  left panel but scaled by $(m_l/m_s)^{1/2}$. The behavior of the peak as a
  function of the quark mass is discussed in the text for both panels.}
\label{fig:chi_l_disc_m}
\end{figure*}

Preliminary unpublished data from the $N_\tau = 8$ simulations is
shown in Fig.~\ref{fig:chi_l_disc_m}.
The left hand figure shows data for $m_l / m_s = 0.2$, $0.1$ and
$0.05$ for the two actions. We find that the height of the peak grows
with decreasing quark mass and is expected to diverge with an exponent 
${1/\delta -1}$ corresponding to the $O(4)$ unversality class of the chiral transition. 

The second noteworthy feature is that the peak is broad and its width
increases with decreasing quark mass.  This is also expected because
in the chiral limit the pions are massless for all values of $T \leq
T_c$ and give rise to a singularity in the ``magnetic'' part of the
free energy, and the chiral susceptibility is expected to diverge as
$m_l^{1/2}$~\cite{karsch09}.  The plot on the right hand side shows
the same data scaled by $(m_l/m_s)^{1/2}$. The peaks in the three data
sets collapse together, consistent with the expected $m_l^{1/2}$
behavior. We do not find simultaneous evidence for the
expected $O(N)$ scaling of the height of the peak mentioned above.
This dual scaling behavior raises the question $-$ what point should
be labeled as $T_c$?  In the absence of a clear peak, we will choose
the right edge to denote the chiral symmetry restoration temperature
$T_c^{\rm chiral}$.

The third feature shown by the data is that the location of the peak
moves to lower temperatures with decreasing quark mass.  For $m_l =
0.1 m_s$ the location is consistent with the $185-195$ MeV range found
earlier, however, the trend suggests a lower value (by $5-10$ MeV) in
the physical quark mass limit~\cite{Cheng:2009zi}. Similarly, the location of the peak in
the asqtad action is shifted with respect to p4 to lower temperatures
at each value of the quark mass by $\sim 5$ MeV.

The bottom line is that we have identified a number of systematic
uncertainties that are each $\sim 5$ MeV and go in the same direction
of lowering the estimate for $T_c$. It is therefore necessary that the
continuum along $ m_l = 0.038$ LCP is taken in order to extract
physical results.

\section{Continuum Limit and HotQCD Plans for the Future}

The results presented in the previous sections have been obtained on
lattices with $N_\tau=6$ and $8$ along a LCP defined by $m_l = 0.1
m_s$. Two extrapolations are needed to obtain physical results. These
are extrapolations in the light quark mass to the physical value $m_l
\approx 0.038$ and the continuum limit $N_\tau \to \infty$ with $a \to
0$. The present data do not allow us to make either extrapolation
reliably.

A number of simulations are in progress to address these two limitations:
\begin{itemize}
\item For both asqtad and p4 actions and for $N_\tau=6$ and $8$
  lattices we are completing simulations at three values of the light
  quark masses, $i.e.$ $m_l/m_s = 0.2, \ 0.1$ and $0.05$. These data
  will allow us to make extrapolations to the physical quark mass and
  the chiral limit.
\item Having established in this study that asqtad and p4 actions give
  consistent results, we have started new asqtad simulations with
  $N_\tau = 12$ and $m_l = 0.05 m_s$. The combined data at $N_\tau =
  6,\ 8$ and $12$ will allow us to take the continuum limit along the
  LCP defined by $m_l = 0.05 m_s$ and extrapolated data at $m_l = 0.038 m_s$.
\item We are exploring the Highly Improved Staggered Quark (HISQ)
  action~\cite{follana07} which has been designed to improve both the
  $O(a^2)$ scaling behavior and reduce taste symmetry violations~\cite{Alexei_hisq09}.
  As mentioned earlier, the uncertainties associated with taste
  symmetry violations are not well understood and could be large.
\end{itemize}

\section{Comparison of $T_c$ with Results from the Wuppertal-Budapest Collaboration}

The Wuppertal-Budapest collaboration (WBC) has recently presented the
following results on the location of the crossover indicating
deconfinement and chiral symmetry restoration:
\begin{eqnarray}
T_c & = & 146(2)(3) \ {\rm MeV} \ \ {\rm Chiral}, \nonumber \\
T_c & = & 170(4) \ {\rm MeV} \ \ \ \ \ \  {\rm Deconf}.
\label{eq:WBresults}
\end{eqnarray}
WBC results include extrapolation to the continuum limit using $N_\tau
= 8, \ 10, 12$ lattices along the $m_l = 0.04 m_s$ LCP and are
therefore labelled ``physical''.  They raise the following two issues
when comparing to the HotQCD results discussed earlier.
\begin{itemize}
\item The WBC estimate for the chiral transition at $146(2)(3)$ MeV is
  about $40$ MeV lower than the HotQCD results.
\item The crossover in the chiral symmetry and deconfinement sectors
  are distinct and separated by $\approx 24$ MeV whereas all HotQCD
  data indicate they are coincident.
\end{itemize}

Of these, we consider the difference in the location of the chiral
transition the more significant issue that needs to be resolved. The
reasons for discounting the difference in the location of
$T_c^{\rm deconf}$ as significant are
\begin{itemize}
\item The best HotQCD data are on $N_\tau =8$ with $m_l = 0.1 m_s$
  lattices. Three systematic effects, each shifting the estimate by
  $\approx 5$ MeV to lower temperatures, have been identified. These
  are (i) the difference between $N_\tau = 6$ and $8$ lattice data;
  (ii) extrapolation in light quark masses and (iii) the difference
  between asqtad and p4 data. A fourth possible shift of $5 $ MeV arises
  from the recent change in the experimental value of $f_K$ which
  resulted in lowering the WBC results by $\sim 5$ MeV~\cite{aoki09}.
  This change affects the continuum extrapolation carried out by the
  WBC and not the HotQCD results presented at fixed $a$ with the scale
  set by $r_0$. All four factors are not necessarily independent,
  however, the point we want to make is that the difference between
  $170(4)$ and $185-195$ MeV can largely be accounted for by a combination of
  these systematic effects.
\item More important, we have presented reasons for why the two
  methods used by the WBC to extract $T_c^{\rm deconf}$, the
  Polyakov loop and strange quark number susceptibility, do not probe
  the singular structure of the theory.  Our contention is that all
  previous results based on these probes do not qualify as precision
  measurements.
\item Probes sensitive to the singular structure (energy and entropy
  density and the light quark number susceptibility) show the two
  transitions to be coincident for all values of the lattice
  parameters simulated by the HotQCD collaboration. It remains to be
  seen if this conclusion survives after extrapolation to the
  continuum limit.
\end{itemize}

The difference in $T_c^{\rm chiral}$ is less significant than it
appears. All three systematic effects in HotQCD calculations discussed
above as well as the change in WBC estimates on using the new
experimental value of $f_K$ are again applicable.  Even though the WBC
find that the five quantities ($r_0$, the new experimental value of
$f_K$, $M_\Omega$, $M_{K^*}$ and $M_\phi$) used to set the lattice
scale give consistent estimates in the continuum limit, there can be
systematic differences when working at a given $a$ as is the case for
the HotQCD data.  Assuming these four $\sim 5$ MeV shifts add, as
data indicate, the HotQCD estimate would be lowered to $\sim 170$ MeV
in the continuum. This, however, still leaves a difference of about
$24$ MeV that needs to be resolved.

The good news is that the HotQCD simulations underway will allow us to
extrapolate to the continuum limit and to the physical light quark
mass and thus shed light on these issues. We hope to have these
results within the next year.

\begin{acknowledgments}
I thank Derek Teaney for inviting me to present the HotQCD results at
this meeting and to participate in exciting discussions on the RHIC
program.

\end{acknowledgments}

\bigskip 

\end{document}